\documentclass[preprint]{aastex}

\usepackage{amsmath}
\usepackage{color}
\usepackage{mathtools}

\shorttitle{Meridional Flow Measurements}
\shortauthors{Jackiewicz et al.}

\newcommand{\id}{ {\rm d} }

\newcommand{\bvec}[1]{ \mbox{\boldmath$#1$} }

\newcommand{\T}{  {\scriptsize{\rm T}}}

\begin{document}

\title{Meridional Flow in the Solar Convection Zone II: \\
           Helioseismic Inversions of GONG Data}

\author{J. Jackiewicz}
\affil{Department of Astronomy, New Mexico State University, Las Cruces, NM 88003}
\email{jasonj@nmsu.edu}

\author{A. Serebryanskiy}
\affil{Ulugh Beg Astronomical Institute, Uzbek Academy of Science, Tashkent 100072, Uzbekistan}

\author{S. Kholikov}
\affil{National Solar Observatory, Tucson, AZ 85719}


\begin{abstract}

Meridional flow is thought to play a very important role in the dynamics of the solar convection zone; however, because of its relatively small amplitude, precisely measuring it poses a  significant challenge. 
Here we present a complete time-distance helioseismic analysis of about two years of ground-based GONG Doppler data to retrieve the meridional circulation profile for modest latitudes, in an attempt to corroborate results from other studies. We use an empirical correction to the travel times due to an unknown center-to-limb systematic effect. The helioseismic inversion procedure is first tested and reasonably validated on artificial data from a large-scale numerical simulation, followed by a test to broadly recover the solar differential rotation found from global seismology.
From GONG data, we measure poleward photospheric flows at all latitudes with properties that are comparable with  earlier studies, and a shallow equatorward flow about $65$\,Mm beneath the surface, in agreement with recent findings from HMI data. No strong evidence of multiple circulation cells in depth nor latitude is found, yet the whole phase space has not yet been explored. Tests of mass flux conservation are then carried out on the inferred GONG and HMI flows and compared to a fiducial numerical baseline from models, and we find that the continuity equation is poorly satisfied.
While the two disparate data sets do give similar results for about the outer $15\%$  of the interior radius, the total inverted circulation pattern appears to be unphysical in terms of mass conservation when interpreted over modest time scales. We can likely attribute this to both the influence of realization noise and  subtle effects in the data and measurement procedure.
\end{abstract}


\section{Introduction}

It is well established that an accurate measurement of solar meridional circulation would be of enormous value as an input to the complex models used to study the solar dynamo and large-scale convection-zone dynamics \citep{charbonneau2014}. Furthermore, debate still continues as to the precise mechanism for generating and sustaining these poleward and equatorward flows and their intricate interplay with differential rotation and thermal structure \citep{rempel2006,miesch2011,dikpati2014}: different parametrizations in numerical models produce drastically different outcomes. A better understanding would also have significant impacts for studies of other stars with surface convection zones \citep{brun2009,featherstone2015}.

Over the past two decades, many observational attempts have been made to quantify this important phenomenon. Methods used have generally included Doppler-shift diagnostics and cross-correlation of features techniques \citep{komm1993,hathaway1996,ulrich2010,hathaway2012} for relatively near-surface studies,  the various tools associated with local helioseismology \citep{giles1997,braun1998, hernandez1999,woodard2012,komm2013,zhao2013,zhao2014}, and even global helioseismology \citep{schad2011,schad2013} to probe the entirety of the convection zone. While a detailed comparison of all of the results from these analyses will not be presented here, it is reasonable to state that no consistent, coherent, nor satisyfing picture of the full circulation profile  has  yet to emerge. In particular,  observational signatures that distinguish between the canonical single, large meridional cells or multiple cells stacked in radius or latitude have not become readily apparent.


In a recent paper \citep[][hereafter Paper I]{kholikov2014},  measurements of travel-time differences of almost two years of ground-based Global Oscillation Network Group (GONG) data  showed evidence of the signature of a rather complex structure of the large-scale flows in the solar convection zone. Here, we add to this rich body of literature and introduce a helioseismic inversion procedure, test it with numerical simulations and ``known'' results, and finally invert the GONG travel times to recover flows. Without any strong  reason to think that our results should significantly differ from others, or will convincingly demonstrate once and for all what the real Sun is doing, we nonetheless find reasonable agreement with  the space-based Helioseismic and Magnetic Imager \citep[HMI, ][]{scherrer2012,schou2012} inferences \citep{zhao2013} in the near-surface after applying a similar systematic correction to the travel times. 

\S~\ref{data} briefly summarizes the GONG measurements and \S~\ref{method} describes the forward and inverse method we apply.  We test the inversion technique  in \S~\ref{test} with artificial data from a numerical model of meridional circulation and using the well-known differential rotation profile.  While we do not discuss at any great length how the results shown here fit into  context of the overall body of literature on this subject mentioned above (this is reserved for an upcoming publication), in \S~\ref{disc} we  present our results, but then seriously question how realistic they are with regards to mass conservation in the convection zone. Conclusions are provided in \S~\ref{conc}.

\section{Data and Measurements}
\label{data}

We use the  measurements that were described in Paper I. To briefly summarize, 652 days of tracked GONG Doppler velocity images were used to construct travel-time differences using the spherical harmonic decomposition technique \citep{kholikov2011}. Skip distances from $3^\circ$ to $47^\circ$ were considered, which correspond approximately to lower turning points of $0.98\,R_\odot$ to $0.70\,R_\odot$ (angular degree $0-300$). After averaging over longitude, the final ``coverage'' of the south minus north (S-N) travel times on the solar disk is within the latitude range of about $\pm75^\circ$ for the shortest distances to about $\pm40^\circ$ latitude for the longest.

To correct for the center-to-limb (CTL) systematic effect as first presented in \citet{zhao2012}, the identical measurement procedure was carried out to obtain east minus west (E-W) travel-time differences. Essentially, we simply rotate the Doppler images by 90 degrees and use the same data-analysis pipeline.

Results were shown in Figures 2 and 3 of Paper I. What we found was that, unsurprisingly, initial S-N travel-time differences gave counterintuitive results, in that the signal became stronger at larger  depths (skip distances). Presumably, as mentioned above, this is due to an unknown artifact that trends in the center-to-limb direction.  After the subtraction of the E-W measurements from the S-N ones, which has been argued will correct for such a trend \citep{zhao2012}, the S-N travel times show interesting behavior that suggests  a signature of the equatorward return flow, and possibly multiple cell structures.  In the inversion analysis presented here we use the E-W corrected S-N travel-time differences from Paper I, and revisit the systematic effect in the discussion section.




\section{Forward and Inverse Methods}
\label{method}

\subsection{Sensitivity Kernels}

We work in the acoustic ray theory \citep[e.g.][]{kosovichev1997}, whereby travel-time differences $\delta\tau$ of waves propagating in opposite directions are assumed to be caused only  by slowly-varying interior flows along a given ray path. The formalism is usually expressed as
\begin{equation}
  \delta\tau = -2\int_\Gamma \frac{\bvec{\mathrm n}\cdot\bvec{u}}{c_{\rm s}^2}\id s,
  \label{dt}
\end{equation}
where $\bvec{u}$ is the flow, $\bvec{\mathrm n}$ is a unit vector tangent to the unperturbed ray path $\Gamma$ and $c_{\rm s}$ is a model solar sound speed profile. Working in a coordinate system where $r$ is the distance from the Sun's center and $\theta$ the latitude,   Eq.~(\ref{dt}) can be recast in the form 
\begin{equation}
  \delta\tau = \int_\Gamma \bvec{K}(\theta,r)\cdot\bvec{u}(\theta,r)\,\id s,
  \label{dtk}
\end{equation}
where $\bvec{K}$ are known as sensitivity kernels. For the study  here, only the horizontal, or meridional, component of the kernels and flows is considered. Ignoring the radial component of the flow is presumably justified since its influence on the travel-time differences is small, although further discussion of this is given in \S\,\ref{disc}. To interpret the GONG travel times, we compute kernels for each measurement, namely for all 45 values of $\Delta$ in the range of $3^\circ$-$47^\circ$, using $3300\,\mu$Hz for  the value for the central frequency.


\subsection{SOLA Inversions}
\label{inv}

We solve Eq.~(\ref{dtk}) by inverting the travel-time differences and obtain the meridional flow as a function of latitude and depth using the subtractive optimally localised averaging, or SOLA \citep{pijpers1994,jackiewicz2012} technique. To simplify the description we introduce the following notation: let $\delta\tau^{\alpha,\beta}$  be the set of measured travel-time differences, $\alpha = \{\Delta_{\rm obs}\}$ the set of skip distances, and $\beta = \{\theta_{\rm obs}\}$ the set of central latitudes.

Let the position of interest at which we would like to estimate the flow speed $u$ in the solar interior, i.e.,  the target location at some depth and latitude, be $(\theta_\T,r_\T)$. The goal of the inversion is to find a  linear combination of the travel times
 \[ u(\theta_\T,r_\T) = \sum_{\alpha,\beta}w_{\alpha,\beta}(\theta_\T,r_\T)\delta\tau^{\alpha,\beta} \]
 with yet to be determined weight functions $w_{\alpha,\beta}$. To derive the weights, consider the cost function
\begin{multline}
\chi(w;\mu) = \int_\theta\int_r  \bigg\rvert\sum_{\alpha,\beta} K^{\alpha,\beta}(\theta,r)w_{\alpha,\beta}(\theta_\T,r_\T)-T(\theta,r;\theta_\T,r_\T))\bigg\rvert^2\id\theta\,\id r\\
+\mu\sum_{\alpha,\beta}\sum_{\alpha',\beta'}w_{\alpha,\beta}(\theta_\T,r_\T)\Lambda_{\alpha,\beta}^{\alpha',\beta'}w_{\alpha',\beta'}(\theta_\T,r_\T),
\label{L2}
\end{multline}
 where $K^{\alpha,\beta}$ are the sensitivity kernels described earlier and $\mu$ is a trade-off parameter that controls  localization and error propagation.  $T(\theta,r; \theta_\T,r_\T)$ is the target function, typically a Gaussian centered at the position of interest that is normalized as 
 \[ \int_{\theta}\int_r T(\theta,r; \theta_T,r_T)\, \id\theta\, \id r = 1. \]
Averaging kernels, seen as the first term in Eq.~(\ref{L2}), are similarly normalized:
\[ \int_{\theta}\int_r\widetilde{K}(\theta,r;\theta_\T,r_\T)\,\id\theta\,\id r =\int_{\theta}\int_r \sum_{\alpha,\beta}w_{\alpha,\beta}(\theta_\T,r_\T)K^{\alpha,\beta}(\theta,r)\,\id\theta\,\id r = 1. \]
With these constraints and an added Lagrange multiplier $\lambda$, the minimization is performed on the the functional
\begin{equation}
  \Im (w;\lambda, \mu) = \chi(w;\mu) + \lambda \left(\int_\theta \int_r \widetilde{K}(\theta,r;\theta_\T,r_\T)\id\theta\,\id r-1\right),
\end{equation}
whereby ones solves the coupled equations
\begin{equation}
  \frac{\partial\Im}{\partial w_{\alpha,\beta}(\theta_\T,r_\T)} = 0, ~~~
   \frac{\partial\Im}{\partial \lambda} = 0.
\end{equation}
The travel-time noise covariance matrix, which is used to compute the variance of the  noise propagated in the inversion (second term in Eq.~\ref{L2}), has the form
 \[ \Lambda^{\alpha,\beta}_{\alpha',\beta'} = 
 {\rm Cov} \biggl[ N^{\alpha,\beta}, N^{\alpha',\beta'} \biggr]. \]
The error in each travel-time measurement, due to random realization noise, is captured in $N$. In our particular analysis here, this  covariance matrix is diagonal, and since the kernels $K_{\alpha,\beta}$ and measurements $\delta\tau$ are normalized by the error in the set of observations, the covariance matrix is simply the identity matrix. 

Given this inverse problem the solution can be found as
 \[ w_{\alpha,\beta}(\theta_\T,r_\T) = \textbf{A}_{\alpha,\beta}^{-1}\textbf{b}_{\alpha,\beta},\]
 where
 \[\textbf{A}_{\alpha,\beta} =  \int_{\theta}\int_r K^{\alpha,\beta}(\theta,r)
 K^{\alpha',\beta'}(\theta,r)\,\id\theta\id r + \mu \Lambda^{\alpha,\beta}_{\alpha',\beta'}, \]
 \[\textbf{b}_{\alpha,\beta} = \int_{\theta}\int_r K^{\alpha,\beta}(\theta,r)T(\theta,r; \theta_T,r_T)\,\id\theta\id r.\]
 When solving the inverse problem with the SOLA method, $\textbf{A}_{\alpha,\beta}$ is often poorly conditioned and its inverse can be found using a single value decomposition (SVD) method, for example. There are two ways to smooth the resulting solution: (i) use two parameters for regularization: $\mu$ for the error covariance matrix, and some threshold paramenter for the eigenvalues to compute the inverted matrix using the SVD algorithm; (ii) increase the values of the target Gaussian sizes in the two relevant dimensions,  FWHM$_{\theta}$ and/or FWHM$_r$. In this work we used different values for all of these parameters to find an optimal set, which  is chosen by inspection of the  so-called L-curves.


\section{Validation Tests}
\label{test}

We provide two examples to  test and validate the procedure described above. In this section we apply helioseismology to  artificial data from a  numerical model of enhanced meridional flow, followed by an inference of the well-studied solar differential rotation profile. We find that within the errors the method  acceptably recovers the known flow profiles.

\subsection{Numerical Simulation}

The artificial data is described in \citet{hartlep2013}.  This simulation has an imposed meridional profile as shown in  the leftmost panel of Fig.~\ref{fig:hartlep_inv}. It has the ``classic'' structure of  poleward  motions in the outer $15\%$ in radius and an equatorward flow starting at depths of 150\,Mm in each hemisphere. To increase the signal-to-noise without running the simulation for an unreasonable amount of time, the authors increased the model flow speed (from its original formulation) by a factor of 36 to give a surface maximum value of $500\,{\rm m\,s^{-1}}$, and a maximum return speed of roughly  $100\,{\rm m\,s^{-1}}$ \citep{hartlep2013}.

\begin{figure}[t]
  \centering
  \includegraphics[width=.8\textwidth]{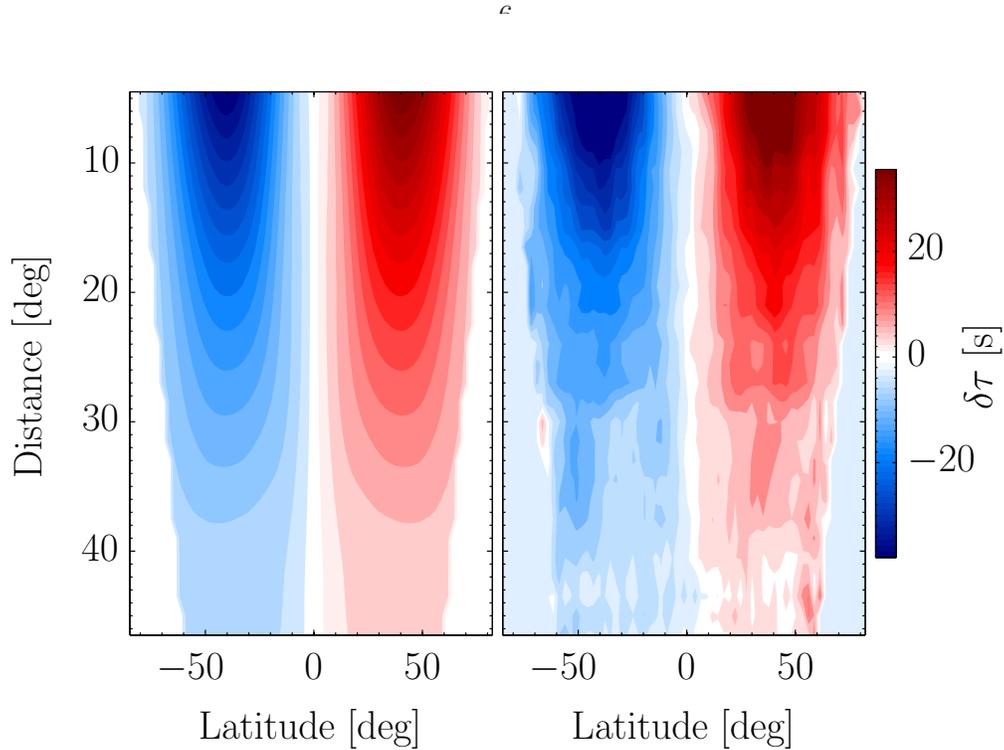}
  \caption{Forward modeled travel-time differences in the S-N sense (left) and measured ones (right) for the simulated meridional flow \citep{hartlep2013} as functions of latitude and skip distance. The color scale is the same in each panel.}
  \label{fig:hartlep_tt}
\end{figure}

Travel-time differences in the south-north  sense are measured for 198 latitudes in the range of $-85.695^\circ$ to  $+85.695^\circ$ and 29 distances in the range of $4.5^\circ - 46.5^\circ$. Flow sensitivity kernels are computed for each distance, and are used to compute forward travel times using Eq.~(\ref{dtk}). A comparison of the two sets of travel times are given in  Fig.~\ref{fig:hartlep_tt}. As expected, forward travel times are smooth (the noise is not modeled). The measured travel times show various structure due to noise, and are furthermore not symmetric about the equator, even though the flow model is. Errors are not shown but rms values are on the order of about $1\,$s at all distances, and are not a strong function of latitude since this is not a real solar data set. These results are comparable to the helioseismic measurements that were  shown in \citet{hartlep2013}.

Both the forward and measured travel times are inverted using the methods described in \S~\ref{inv} to  recover the strong flows in the model. An optimal set of weights is computed and used to combine each set of travel times. The results of the inversions are shown in Fig.~\ref{fig:hartlep_inv}. Excellent agreement is seen for the case using the forward travel times (middle panel), thus providing a convincing validation of the inversion method and codes.  Due to somewhat noisy travel times, the flows shown in the rightmost panel of  Fig.~\ref{fig:hartlep_inv} exhibit features that are not completely consistent with  the model. The return flow is detected; however, it is  truncated in its latitudinal extent, most noticeably near the deep equatorial region.  The inversion errors in the region deeper than $0.85R_\odot$ have an rms of  $45\,{\rm m\,s^{-1}}$, while those shallower than that are about $20\,{\rm m\,s^{-1}}$. 

To some degree, and without an obvious  alternative on hand, this exercise justifies the ray approximation for our purposes  in modeling travel-time differences related to large-scale meridional circulation. It also allows us to find optimal inversion parameters that will be applied to GONG data in \S\,\ref{disc}.

\begin{figure}[t]
  \centering
  \includegraphics[width=.8\textwidth]{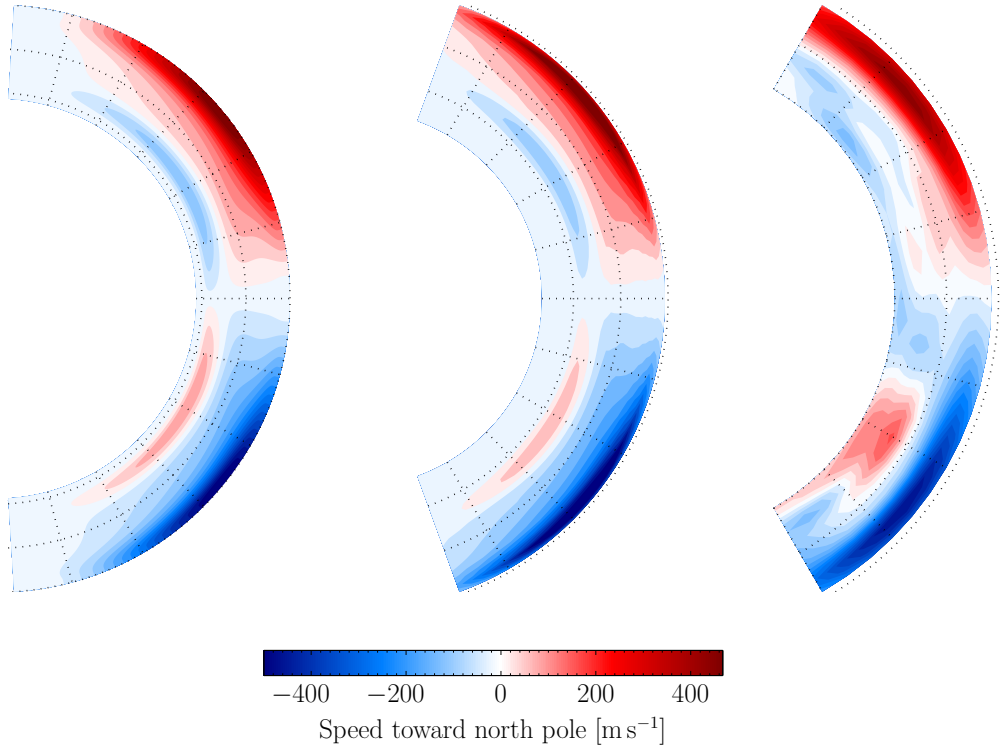}
  \caption{Simulation and inversion results. Shown are meridional cuts of the  flow model (left), the inversion of forward travel-time differences (middle), and inversion of measured travel-time differences (right). For reference, dotted radial lines are plotted every $15^\circ$ and dotted concentric lines are shown for depths $r = [0.7, 0.85, 1.0]R_\odot$. Not shown is the model radial flow profile. All plots use the same color scale, but note the different depth and latitude ranges.}
  \label{fig:hartlep_inv}
\end{figure}

\subsection{Recovering Differential Rotation with Time-Distance Helioseismology}

The  solar differential rotation obtained from global-mode seismology (frequency splittings) is a very robust measurement \citep{schou1998,howe2009b}, and serves as a useful test for other methods. Local helioseismology has also shown  success  in this realm \citep[e.g.][]{giles1998,basu1999}. We attempt a validation of the specific methods described here using real solar data from GONG. 

 In addition to the E-W measurements discussed in \S\,\ref{data} of tracked data, we also made  travel-time measurements of untracked data that preserve rotation information.  The geometry is such that cross correlations for a given latitude and separation distance are computed from point-to-arcs, where the point is always on the central meridian and the arcs lie in one of the hemispheres centered on that latitude. The westward signal (prograde) is subtracted from the eastward signal (retrograde), leaving behind the expected influence of rotation that is  subsequently free from the CTL effect described earlier. In other words, since a purely symmetric CTL effect has the same sign in each hemisphere with this measurement geometry, upon subtraction it is canceled out.  These calculations are then done for many distances and latitudes, for about 360 days of data.

\begin{figure}[t]
  \centering
  \includegraphics[width=\textwidth]{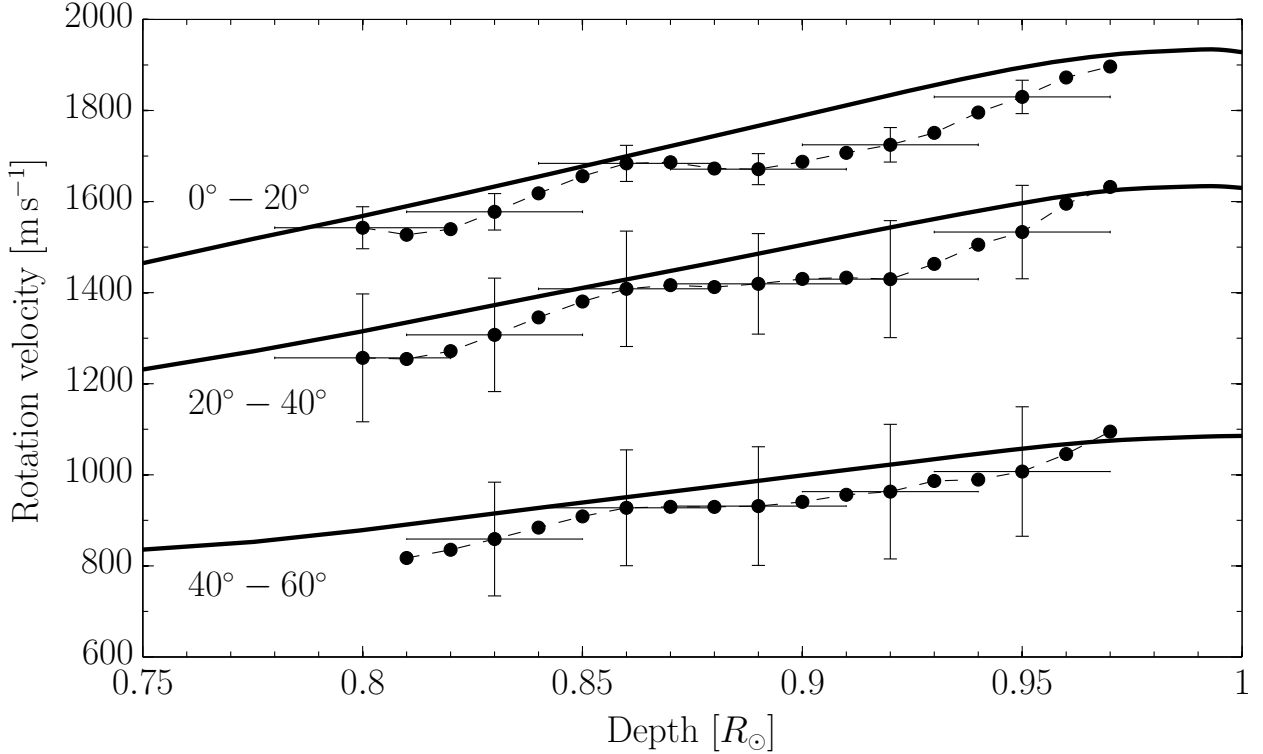}
  \caption{Comparison of the solar (differential) rotation velocity estimated from inverting  travel-time differences  (dashed lines with circles) with global helioseismic results (solid lines). Averages are computed over three latitude bins as labeled. Inversion errors in inferred velocity and uncertainties from the target function width are shown every three depth points for clarity.}
  \label{fig:rotation}
\end{figure}

To invert for solar rotation we use the following set of parameters: the location of the target function in depth is set from $0.8R_\odot$ to $1.0R_\odot$ with steps of $0.01R_\odot$ and in latitude from $-60^\circ$ to $+60^\circ$ with steps of $2.5^\circ$; the width of the target function in depth and latitude is  FWHM$_r=0.04R_\odot$ and FWHM$_\theta=5^\circ$, respectively. To make a more meaningful comparison, the global helioseismic results are smoothed to our inversion resolution by convolving with the target function used at each depth. Finally, both sets of results are averaged over three latitude bins from $0^\circ-20^\circ$, $20^\circ-40^\circ$, and $40^\circ-60^\circ$. 

The comparison with the time-averaged rotation profile from global analysis of about 1000 days of HMI data \citep{howe2011} is shown in Fig.~\ref{fig:rotation}. Within the uncertainties of our local-helioseismic inversion, the results are in agreement over most depths and latitudes except for the near-surface region around the equator. There are many possible contributions to any disagreement in general, including different instruments and non-overlapping time series. Despite some discrepancies, this provides more confidence that large-scale flows, albeit strong ones in this case, are somewhat accurately attainable with these tools.


\section{Results and Discussion}
\label{disc}

\subsection{Meridional Flows from GONG Data}

\begin{figure}[t]
  \centering
  \includegraphics[width=\textwidth]{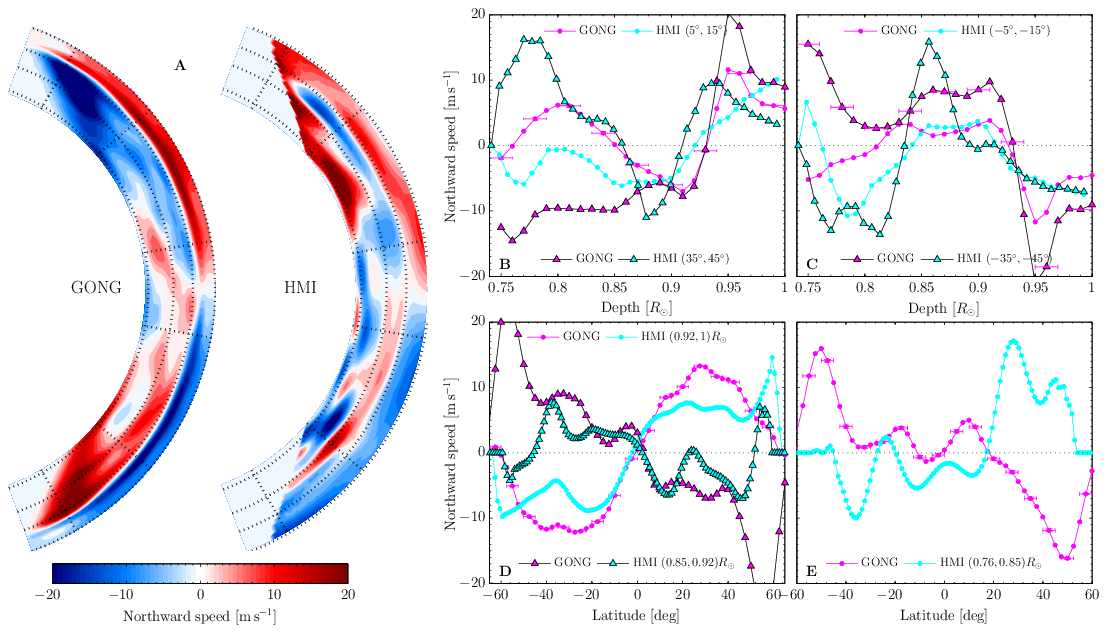}
  \caption{Comparison of meridional flows measured from GONG and HMI data. Panel A shows cross-sectional views of the meridional flow profiles within latitudes of $\pm70^\circ$ and depths above $0.74R_\odot$.  The dotted lines on each image are plotted at depths $r=(0.76, 0.85, 0.92, 1.0)R_\odot$ and at latitudes $\theta=\pm(10^\circ, 40^\circ, 60^\circ)$, for reference in the line plots on the right. Panels B and C show velocities as a function of depth averaged over $10^\circ$ latitude bands, as labeled in the legend, for the nothern and southern hemispheres. Panels D and E show velocities as a function of latitude averaged over the depth ranges shown in each legend. Amplitude error and radial resolution bars, shown only for GONG every several points, are similar for HMI.}
  \label{fig:gong_hmi}
\end{figure}

The CTL-corrected S-N travel times measured from GONG data and introduced in \S\,\ref{data}, were inverted for flows  and are shown in Fig.~\ref{fig:gong_hmi}. Also provided are the HMI results from \citet{zhao2013} that were kindly provided by J.~Zhao for comparison. There are several noteworthy remarks:
\begin{itemize}\itemsep 0cm
\item GONG and HMI show very good agreement  regarding the location of the change of sign from surface poleward flows to equatorward flows at about $r=0.91R_\odot$. This is consistent in both hemispheres, as can be seen in panels A, B, and C  of Fig.~\ref{fig:gong_hmi}. 

\item The peak near-surface poleward flows occur at about $30^\circ$ latitude in each hemisphere for both data sets (panel D), with amplitudes in the $15-20\,{\rm m\,s^{-1}}$ range.

\item The shallow return flow peaks at similar latitudes with amplitude around $5-10\,{\rm m\,s^{-1}}$.

\item Unlike HMI, GONG only shows weak evidence of a second ``cell'' at low latitudes within $15^\circ$ of the equator, a feature  we suspect is spurious. The equatorward circulation at the bottom of the detection region is consistent with recent predictions from flux-transport models \citep{hazra2014}.

\item Any features in high-latitude regions  ($\geq 55^\circ$) are suspicious, particular the strong equatorward flows in the GONG results.

\item  The results of HMI and GONG are basically anticorrelated at the deepest layers, yet both interestingly cross zero at the same latitude ($20^\circ$, panel E).

\item  We do not observe any evidence of multiple cells in latitude at any depths, unlike the \citet{schad2013} study.

\end{itemize}

There are also several (of likely many) caveats to keep in mind:
\begin{itemize}\itemsep 0cm

\item Due to the use of  long time series, the random errors propagated through the inversion are small. However, we caution that uncertainties due to known and unknown systematics \citep{duvall2009} are likely much larger, and would particularly affect the results deep in the convection zone and at high latitudes.

\item That the near-surface meridional flow is not strongest at the very topmost boundary is due to the fact that the smallest skip distance is still rather large (using medium-$\ell$ data), and that there is a level of smoothing in depth  due to averaging kernels in the inversion.

\item To do a more detailed comparison, several independent sets of measurements are needed, for example, analysis of separate four-year data sets from 1996-2014 using GONG, MDI, and HMI would be ideal . Efforts are currently underway to do this.
\end{itemize}



Given these methods and results, one can certainly argue that the ad-hoc correction of the CTL systematic is problematic, as the signal one is searching for in the large-distance (deep) travel-time differences is of  order one second.  Subtracting sets of large numbers to reveal  this signal is not a recommended practice, and even though it was convincingly demonstrated in \citet{zhao2012b} that this correction brought multiple HMI observables into agreement, further justification is  certainly needed. One could imagine that the efficacy of the  systematic effect removal should be shown first by  testing it to recover well-known features, such as the solar rotation. Unfortunately, the geometry of the E-W measurements  removes the effect in each hemisphere as mentioned earlier.

\subsection{Mass Conservation in the Convection Zone}

\begin{figure}[h]
  \centering
  \includegraphics[width=\textwidth]{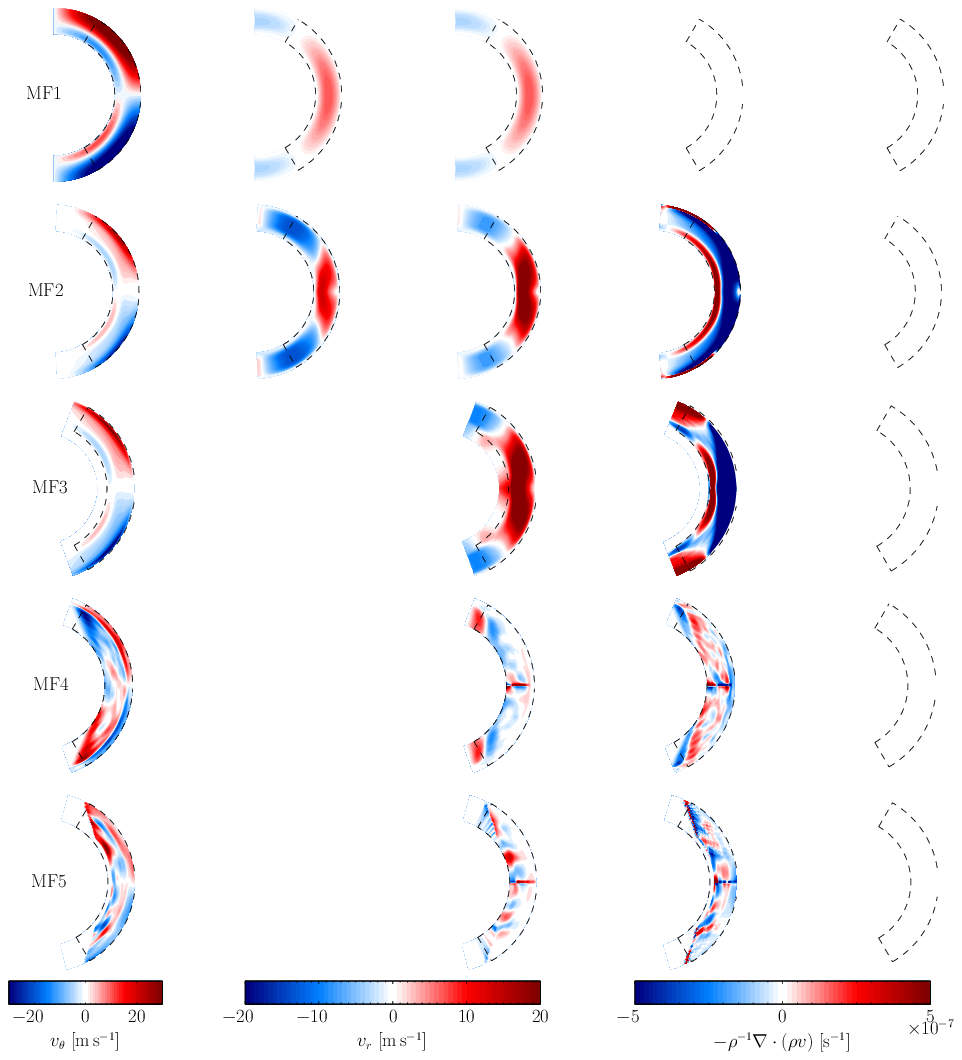}
  \caption{Tests of continuity equation for the 5 meridional flow models (MF\#) described in the text. The columns show the latitudinal speed (column 1), the radial speeds (columns 2-3), and  the continuity computation (columns 4-5). In each panel the dashed reference line encloses the $\pm 60^\circ$ and $(0.7-1.0)R_\odot$ interior region. The large flows associated with the \citet{hartlep2013} model (MF2 and MF3) have been scaled down for plotting to achieve similar values of the other flows (see Table~\ref{tab:mc} for unscaled numbers).}
  \label{fig:matrix}
\end{figure}

It is important to consider if the resulting steady flow fields from the inversions are physically plausible. What is not typically shown in the relevant helioseismic literature is  how derived flows satisfy our notions of a convection zone where  material does not enter or leave to any large degree. Although it is possible to use mass conservation  as an additional constraint directly in the inversion equations \citep[e.g.][]{giles2000}, we have not implemented that. At zeroth-order, we have already ignored inverting for the radial flow.  Here we discuss if it even matters to do detailed comparisons of meridional circulation solutions since there may be larger issues to first identify.

We start with the continuity equation
\begin{equation}
  \frac{\partial\rho}{\partial t} + \bvec{\nabla}\cdot (\rho\bvec{v}) = 0
  \label{mc}
\end{equation}
for some velocity field $\bvec{v}$. Ignoring any latitudinal or azimuthal dependence of the density $\rho$, the expression for mass conservation in  spherical coordinates can be expressed as
\begin{equation}
\bvec{\nabla}_r\cdot(\rho\bvec{v}_r) + \bvec{\nabla}_\theta\cdot(\rho\bvec{v}_\theta) =  - \frac{\partial\rho}{\partial t}, 
\end{equation}
or,
\begin{equation}
 \frac{1}{\rho}v_r\frac{\partial\rho}{\partial r} + \frac{1}{r^2}\frac{\partial}{\partial r}(r^2\, v_r) + \frac{1}{r\sin\theta}\frac{\partial}{\partial\theta}(\sin\theta\, v_\theta)  = - \frac{1}{\rho}\frac{\partial\rho}{\partial t},
\label{mc2}
\end{equation}
where $\theta$ is latitude. Since it is not expected that the Sun's density profile is changing over moderate time scales, the degree to which the total divergence of the mass flux $\rho\bvec{v}$ is zero is a good indication of satisfying continuity.   In what follows we consider applying this equation to five meridional flow profiles. Since three of the profiles described below are from inversions that do not have a radial velocity component,  we  compute it from solving the continuity equation \textit{requiring} mass conservation (to some numerical precision). The solution is found using a simple fourth-order Runge-Kutta technique that initializes the radial flow to be zero at the surface boundary point ($r\approx 1.0R_\odot$) and integrates from that point inwards toward the deepest part of the convection zone considered. Once determined, we may then ask if this flow is indeed reasonable.

\begin{table}
\centering
\caption{Continuity parameters for five meridional flow profiles.}
\label{tab:mc}
\small
\begin{tabular*}{\textwidth}{@{\extracolsep{\fill}}l l l r r  r r r}
\tableline\tableline
  Row\# & Quantity & Unit &MF1 & MF2 & MF3 & MF4 & MF5  \\\hline
1 & $v_\theta^{\rm rms}$ & ${\rm m\,s^{-1}}$ & ${\rm 1.79e+01}$ &  ${\rm 1.87e+02}$ &  ${\rm 1.75e+02}$ &  ${\rm 9.45e+00}$ &  ${\rm 6.77e+00}$ \\ 
2 & $v_r^{\rm rms}$ & ${\rm m\,s^{-1}}$   & ${\rm 3.10e+00}$ &  ${\rm 1.43e+01}$ &  ${\rm 0.00e+00}$ &  ${\rm 0.00e+00}$ &  ${\rm 0.00e+00}$ \\ 
3 & $\overline{v}_r^{\rm rms}$ & ${\rm m\,s^{-1}}$ & ${\rm 3.10e+00}$ &  ${\rm 1.84e+01}$ &  ${\rm 2.81e+01}$ &  ${\rm 4.24e+00}$ &  ${\rm 2.99e+00}$ \\ 
4 & $\sum{\rm div}\,\rho v_\theta$ & -  & ${\rm 3.30e-05}$ &  ${\rm 2.32e-04}$ &  ${\rm 4.75e-04}$ &  ${\rm -1.73e-05}$ &  ${\rm 2.82e-05}$ \\ 
5 & $\sum{\rm div}\,\rho v_r$ & -  & ${\rm -3.30e-05}$ &  ${\rm 1.02e-04}$ &  ${\rm 0.00e+00}$ &  ${\rm 0.00e+00}$ &  ${\rm 0.00e+00}$ \\ 
6 & $\sum({\rm div}\,\rho v_\theta + {\rm div}\,\rho v_r)$ & - & ${\rm -2.63e-09}$ &  ${\rm 3.34e-04}$ &  ${\rm 4.75e-04}$ &  ${\rm -1.73e-05}$ &  ${\rm 2.82e-05}$ \\ 
7 & $\sum{\rm div}\,\rho \overline{v}_r$ & - & ${\rm -3.30e-05}$ &  ${\rm -2.32e-04}$ &  ${\rm -4.75e-04}$ &  ${\rm 1.73e-05}$ &  ${\rm -2.82e-05}$ \\ 
8 & $\sum({\rm div}\,\rho v_\theta + {\rm div}\,\rho \overline{v}_r)$ & - & ${\rm 2.23e-09}$ &  ${\rm 3.69e-08}$ &  ${\rm 4.90e-08}$ &  ${\rm 1.83e-09}$ &  ${\rm 2.49e-09}$ \\ 
9 & $\frac{\sum({\rm div}\,\rho v_\theta + {\rm div}\,\rho v_r)}{\sum{\rm div}\,\rho v_\theta}$ & - & ${\rm -7.97e-05}$ &  ${\rm 1.44e+00}$ &  ${\rm 1.00e+00}$ &  ${\rm 1.00e+00}$ &  ${\rm 1.00e+00}$ \\ 
10 & $\frac{\sum({\rm div}\,\rho v_\theta + {\rm div}\,\rho \overline{v}_r)}{\sum{\rm div}\,\rho v_\theta}$ & - & ${\rm 6.77e-05}$ &  ${\rm 1.59e-04}$ &  ${\rm 1.03e-04}$ &  ${\rm -1.06e-04}$ &  ${\rm 8.81e-05}$ \\ 
11 & $\langle-\rho(\partial\rho/\partial t)^{-1}\rangle$ & day & ${\rm -3.36e+06}$ &  ${\rm 2.51e+01}$ &  ${\rm 3.05e+01}$ &  ${\rm -1.82e+02}$ &  ${\rm 2.89e+02}$ \\ 
12 & $\langle-\rho(\partial\rho/\partial t)^{-1}\rangle_{\rm scaled}$ & year & ${\rm -4.57e+09}$ &  ${\rm 3.41e+04}$ &  ${\rm 4.16e+04}$ &  ${\rm -2.48e+05}$ &  ${\rm 3.93e+05}$ \\ 
\hline
\end{tabular*}
\tablecomments{All quantities are computed in the truncated radial region $r=0.75 - 0.995\,R_\odot$. The quantities in rows 11-12 are computed from the divergence of the mass flux using $v_r$, not $\overline{v}_r$, and $\langle\ldots\rangle$ represents the median of the values in the truncated region. Physical units are only provided for relevant quantities.}
\end{table}


\subsubsection{Fiducial model: MF1}
To test such computations and to establish a numerical baseline, we first utilize a model of meridional circulation that by construction analytically satisfies Eq.~(\ref{mc}). Such models are common in the literature, and we choose the one originally described in \citet{vanb1988}, which has the ``canonical'' single-cell deep return flow, and uses a  power law density profile. We  compute this model, denoted ``MF1,'' and run it through the continuity equation to determine the numerical precision to which it is satisfied. The latitudinal and radial flows are shown in columns 1 and 2 for model MF1 in Fig.~\ref{fig:matrix}.  The sum of the horizontal and radial mass flux divergences scaled by the inverse density, i.e., an estimate of $\rho^{-1}(\partial\rho/\partial t)$, is shown in the 4th column.  

As a  test of the initial-value solver, we also compute the radial flow obtained only from  (the same) $v_\theta$. This will be denoted  $\overline{v}_r$ in each case hereafter. Column 3 of  Fig.~\ref{fig:matrix} shows this derived quantity, and we  see that it matches very well the original $v_r$ (column 2). For completeness, we then recompute the continuity equation given the new $\overline{v}_r$ and original  $v_\theta$, shown in column 5. It is easy to see by eye that this model satisfies continuity, as it must by construction.

This template is then followed for 4 more cases below, where all profiles are computed on (or interpolated onto) a grid of 350 points in depth and latitude in the convection zone (experiments with more points were done but do not qualitatively change the conclusions). In addition to Fig.~\ref{fig:matrix}, Table~\ref{tab:mc} shows the quantitative values for a more detailed view of these calculations. Keeping with the example for MF1, the 6th row of the table, which is the sum over the mass-flux divergence, gives the numerical precision we anticipate in this best case, $\sim 10^{-9}$ in arbitrary units. The 8th row gives the same quantity using the recomputed radial velocity, which in this case is very similar.  Rows 9-10 are  important  for this discussion since they show the relative magnitude of the numerical integral of the two terms in the equation with respect to one of the terms. 

Row 11  of  Table~\ref{tab:mc} is the  inverse of the RHS of the density-scaled continuity equation, which gives units of  time. One plausible interpretation of this is that it represents the timescale over which it would take to drain (positive value, $\partial\rho/\partial t<0$) or fill (negative value,  $\partial\rho/\partial t>0$)  a ``typical'' region of the convection zone of that model, with mass. This number is  computed from the median of all the values (column 4 of Fig.~\ref{fig:matrix})  throughout most of the convective domain. So it represents an approximate measure of the lifetime for the bulk region of the convection zone. The MF1 model sets  the baseline for the numerics at about 10\,kyr. 

Another appealing way to think about this is if we take the MF1 model as ground truth (in a numerical sense, by construction) to represent perfect mass conservation in the convection zone over about $t_\odot=4.57\,$Gyr, we find a scaling factor ($\sim 10^5$). This scaling can be applied to the subsequent MF profiles to understand the calculations in terms of  the solar lifetime. This timescale is shown in row 12 for all cases.

\subsubsection{Profiles M2-M5}

We next consider the \citet{hartlep2013} flow model, ``MF2,'' employing  the standard Model S density profile \citep{jcd1996} for the computations. As can be seen in the 4th column of  Fig.~\ref{fig:matrix}  for MF2, the mass is conserved very poorly. We speculate that this is somehow due to the fact that the flows were increased by a significant amount from the original model to provide a detectable signal, as discussed earlier.  Computing the new $\overline{v}_r$ (column 3) to conserve mass improves this (column 5), but requires strong radial flows, with an rms value  about  30\% larger than the original value (Table~\ref{tab:mc}, row 3). Here, the typical region in the convection zone lost its mass over a few tens of thousands of years.

``MF3''  is  the inverted profile of \citet{hartlep2013} shown earlier in Fig.~\ref{fig:hartlep_inv}. The 2nd column of   Fig.~\ref{fig:matrix} is empty as the radial flow is not inverted for. Thus one expects the mass conservation to be poor, and  the 4th column verifies this. Interestingly, it's at the same order-of-magnitude as the original profile itself, even withouta $v_r$.  The new  $\overline{v}_r$, shown in the 3rd column,  serves to help satisfy continuity rather well (column 5).

``MF4'' and ``MF5'' are   the GONG and HMI inversion results, respectively, shown earlier in  Fig.~\ref{fig:gong_hmi}. The radial flows needed to satisfy continuity at a significant level are strong, with  maximal values in concentrated regions  of about $100\,{\rm m\,s^{-1}}$, yet with much smaller rms values.  They also display a complex pattern of downflows to upflows in latitude, that differ quite a bit from each other.  In terms of the draining/filling of the convection zone, both inverted solutions are on the order of $10^5$\,year. It is somewhat surprising that even with no radial flows the inverted meridional profiles from GONG and HMI conserve mass a bit better than the simulated data, but are still about 4 orders-of-magnitude worse than the fiducial case.

What all of this is suggesting is that the flows we are recovering in our inversions can be interpreted as being unphysical if our basic understanding of the solar convection zone is correct. Stated differently, the retrieved circulation cannot be the whole picture, since we find that the radial flows (which were are not inverted for) necessary to satisfy continuity are significant. They also differ dramatically from the ``standard'' picture as in MF1. If these radial flows do exist as required to conserve mass, then they should be measureable. Furthermore, the flows predict that the convection zone would be drained of mass on much shorter timescales than solar evolution, which we are quite certain is not the case. While it is difficult to put too much emphasis on the precise numbers here due to numerical inacuracies, in a relative sense this simple exercise suggests that the helioseismic results should be carefully explored further.


\section{Conclusions}
\label{conc}

Inversions of GONG S-N  travel-time differences that have been ad-hoc corrected for an unknown center-to-limb systematic effect show the canonical surface poleward meridional flow, below which exists  a shallow equatorward component beginning at about 65\,Mm beneath the surface. This is similar to what is seen from HMI data, yet disagreement appears deeper down when  a second poleward component is found only from HMI. These results may indicate, somewhat curiously, that at some level a similar CTL systematic is present in both the ground-based and space-based data, which is not an obvious conclusion \textit{a priori}. 

Irregardless of systematics, convection-zone mass-conservation arguments appear to demonstrate that the inverted flows are problematic, presumably in both  magnitude  and structure. This is difficult to reconcile, and given the similarities in GONG and HMI results, one obvious question is why is there relatively good agreement if they are both incorrect at some level?  We note  that other  recent studies using  local helioseismology for measuring smaller-scale  flows  have been more challenging than orginally anticipated \citep{degrave2014,hanasoge2014,svanda2015}, though there are promising directions \citep[e.g.][]{woodard2014}.

Other improvements are possible, such as the development and use of finite-frequency sensitivity kernels, which could also provide a way to potentially measure the radial contribution of the circulation. Noise covariances could be more accurately determined, and each day there are more and more data that can be analyzed to help reduce uncertainties. Sources of systematics are also being explored. Unlike rotation, meridional flows are expected to be a much weaker signal and it is not surprising that they have been so difficult to measure accurately despite very hard work in the past several decades. Thus, as demonstrated here it is critical to have independent measurements from different groups to corroborate any inferred circulation patterns.




\acknowledgements
We gratefully thank J.~Zhao for providing his inversion results from HMI data, T.~Hartlep for providing his meridional flow simulation data, and an anonymous referee for an excellent numerical suggestion.  This material is based upon work supported by the National Science Foundation under Grant Number 1351311. A.S. acknowledges  support from grant FA-F02-F028 of the Uzbek Academy of Sciences. S.K. was supported by NASA's Heliophysics Grand Challenges Research grant 13-GCR1-2-0036.  This work utilized data obtained by the Global Oscillation Network Group (GONG) program, managed by the National Solar Observatory, which is operated by AURA, Inc.~under a cooperative agreement with the National Science Foundation.

\bibliographystyle{apj}
\bibliography{myrefs}

\end{document}